\documentstyle[preprint,aps]{revtex}
\begin{document}
\draft
\newcommand{\be}{\begin{equation}}
\newcommand{\ee}{\end{equation}}
\newcommand{\bea}{\begin{eqnarray}}
\newcommand{\eea}{\end{eqnarray}}
\title{Critical Exponents of the Four-State Potts Model}
\author{Richard J. Creswick and Seung-Yeon Kim}
\address{Department of Physics and Astronomy,\\
University of South Carolina, Columbia, SC 29208}
\date{\today}
\maketitle
\begin{abstract}
The critical exponents of the four-state Potts model are directly derived
from the exact expressions for the latent heat, the spontaneous
magnetization, and the correlation length at the transition
temperature of the model.
\end{abstract}    

The q-state Potts model\cite{po,wb} is a generalization of the Ising model
that is the two-state Potts model.
Although the Potts model has not been solved exactly,
there have been several exact results at the critical point
for this model in two dimensions.
In 1952 Potts\cite{po} conjectured the exact critical temperatures 
of his model on the square lattice for all q
by a Kramers-Wannier\cite{kw} type duality argument.
In 1971 Temperley and Lieb\cite{tl} showed that the Potts model
can be expressed as a staggered six-vertex model.
Following the equivalence\cite{n1} between the Potts model
and a staggered six-vertex model, in 1973 Baxter\cite{b1}
calculated the free energy of the Potts model at the critical temperature,
and showed that the model has a continuous phase transition for $q\le4$,
and has a first-order phase transition (i.e. has latent heat)
for $q>4$. In 1979 den Nijs\cite{d1} conjectured the thermal scaling 
exponent for $q\le4$ by considering relation between the eight-vertex
model and the Potts model. In 1980 Nienhuis {\it et al.}\cite{nr} 
and Pearson\cite{pe} conjectured independently the magnetic scaling
exponent for $q\le4$ from numerical results.
In 1981 Black and Emery\cite{be} showed the den Nijs conjecture
to be asymptotically exact by using 
the Coulomb-gas representation\cite{ni}
of the Potts model and renormalization-group methods. 
In 1982 Baxter\cite{b2} calculated the spontaneous magnetization of 
the model at the transition point for $q>4$. 
In 1983 den Nijs\cite{d2} verified a conjecture for the magnetic
scaling exponent for $q\le4$ from the scaling behavior 
of the correlation function in the Coulomb-gas representation.
In 1984 Dotsenko\cite{do}
again verified the conjectures for the thermal and magnetic
scaling exponents for $q\le4$ using conformal field theory.
Recently Buffernoir and Wallon\cite{bw} obtained an exact expression
for the correlation length of the Potts model at the critical temperature
for $q>4$ by using Temperley-Lieb algebra\cite{n1}
and a Bethe ansatz\cite{bt}.   
In this paper we derive the critical exponents of the four-state 
Potts model directly from the three main exact results of the
Potts model which are Baxter's calculation of the latent heat 
and the spontaneous
magnetization and Buffernoir and Wallon's calculation of the
correlation length. 
 
The Hamiltonian for the q-state Potts model on the isotropic
square lattice is
\be
{\cal H}=-J\sum_{<i,j>}\delta(\sigma_i,\sigma_j),
\ee
where
$$\delta(\sigma_i,\sigma_j)=\cases{1,&if $\sigma_i=\sigma_j$,\cr
                           0,&if $\sigma_i\ne \sigma_j$,\cr}$$
and $\sigma_i,\sigma_j=1,2,...,q$.  
The latent heat at the critical temperature $T_c$ is given by\cite{b1}
\be
L=2J(1+q^{-{1 \over 2}}) \tanh{\theta \over 2} \prod^\infty_{n=1}
(\tanh n\theta)^2, q > 4,
\ee   
where $\theta$ is defined by
$$2\cosh \theta = q^{1 \over 2}. $$
The zero-field (spontaneous) magnetization at $T_c$ is\cite{b2}
\be
M_0 = \prod^\infty_{n=1} {{1-x^{2n-1}}\over {1+x^{2n}}} , q > 4,
\ee
where $x=e^{-2\theta}$.
The correlation length, $\xi$, at $T_c$ is given by\cite{bw}
\be
\xi^{-1} = 2\ln {\cosh{3\over2}v\over\cosh{1\over2}v} +
4\sum^\infty_{n=1}{(-1)^n\over n}e^{-2nv}(\sinh nv)(\tanh2nv), q > 4,
\ee
where $v$ is defined by
$$2\cosh v=(2+q^{1\over2})^{1\over2}.$$
The behaviors of the latent heat, the zero-field magnetization,
and the correlation length near the limit $q=4$ are expressed by
the following:
\be
L\sim 2\pi J(1+q^{-{1\over2}})
e^{-{\pi^2\over2}(q-4)^{-{1\over2}}}, 
\ee
\be
M_0\sim 2e^{-{\pi^2\over8}(q-4)^{-{1\over2}}},
\ee
and
\be
\xi^{-1}\sim {8\over\sqrt{2}}e^{-\pi^2 (q-4)^{-{1\over2}}}.
\ee
All three expressions have the same limiting behavior.   

In the spirit of scaling theory, the singular behavior in the latent heat
and the spontaneous magnetization can be expressed in terms of
the correlation length:    
\be
L\sim\xi^{-{1\over2}},
\ee
and
\be
M_0\sim\xi^{-{1\over8}}.   
\ee
Note that in (8) the latent heat and in (9) the spontaneous
magnetization vanish as $q-4\to0^+$.
The four-state Potts model can be considered the critical end point
of a sequence of models ($q>4$) with finite latent heats
and finite spontaneous magnetizations. 
From standard scaling theory\cite{cf}, the internal energy per site is
$$e(t,h=0)=\xi^{-d+{1\over\nu}}f_t(t\xi^{1\over\nu},h=0),$$
where the subscript, $t$, means differentiation with respect to
the reduced temperature. The latent heat is given by
$$L=\xi^{-d+{1\over\nu}}\Delta f_t,$$
where $\Delta f_t=f_t(0^+,0)-f_t(0^-,0)$.
Near the second-order transition point, i.e., 
near the limit $q=4$, 
we have
\be
L\sim\xi^{-d+{1\over\nu}}.
\ee
Comparing (8) and (10), we obtain
\be
d-{1\over\nu}={1\over2}.
\ee
Similarly, for the spontaneous magnetization per spin
\bea
M_0 &=& \xi^{-d+y_h}f_h(0,0)\cr
&=& \xi^{-{\beta\over\nu}}f_h(0,0).
\eea
Comparing this with (9) we see that
\be
{\beta\over\nu}={1\over8}.
\ee
For $d=2$, from (11) and (13), the exact values for the scaling and critical
exponents of the four-state Potts model are
$y_t={3\over2}, y_h={15\over8}, \alpha={2\over3}, \beta={1\over12},
\gamma={7\over6}, \delta=15, \nu={2\over3}$, and 
$\eta={1\over4}$
in agreement with values\cite{n2} derived for $q\le4$
from the Coulomb-gas representation\cite{be,ni,d2}
and conformal field theory\cite{do}.    

The critical properties of the four-state Potts model
have been studied extensively as the limiting case 
of a sequence ($q\le4$) of models with continuous phase transitions.
As is often the case, the limit of such a sequence, $q=4$,
exhibits strong corrections to scaling.
It is, therefore, of interest to approach this problem
from the opposite side, and regard the four-state Potts model 
as the limit of a sequence of models ($q>4$) with a discontinuous,
or first-order, transition.
As $q\to4^+$, the latent heat and spontaneous magnetization
at $T_c$ vanish, and the correlation length diverges.  
We have shown that by applying simple scaling arguments 
to exact calculations of $L, M_0$, and $\xi$ at $T_c$, 
one can derive the exact critical exponents and that 
they agree with those obtained for $q\le4$.

\end{document}